\documentclass[twocolumn,showpacs,preprintnumbers,amsmath,amssymb,aps]{revtex4}
\usepackage{graphicx}
\usepackage{epsfig}
\def\lsim{\raise0.3ex\hbox{$\;<$\kern-0.75em\raise-1.1ex\hbox{$\sim\;$}}}
\def\gsim{\raise0.3ex\hbox{$\;>$\kern-0.75em\raise-1.1ex\hbox{$\sim\;$}}}
\newcommand{\be}{\begin{eqnarray}}
\newcommand{\ee}{\end{eqnarray}}

\def\bea{\begin{eqnarray}}
\def\eea{\end{eqnarray}}

\begin{document}
\preprint{SHEP-11-19}

\title{LHC Signals of a Heavy CP-even Higgs Boson in the NMSSM \\
via Decays into a $Z$ and a Light CP-odd Higgs State}
\author{M. M. Almarashi and S. Moretti}
\affiliation{School of Physics and Astronomy, University of Southampton,
Highfield, Southampton SO17 1BJ, UK.}
\date{\today}

\begin{abstract}
We study the $Za_1$ decay mode of a heavy CP-even Higgs boson of the NMSSM, $h_2$,
where $a_1$ is the lightest CP-odd Higgs state of this scenario,
the former produced in association with a bottom-antibottom pair, and find that, despite small event rates, 
a significant (in fact essentially background free) signal should be extractable at the LHC 
with very high luminosity,
so long that a $Z\to jj$ (where $j$ represents a jet) 
and $a_1\to \tau^+\tau^-$ final state is exploited, in presence of $b$-tagging.
\end{abstract}

\maketitle
%

The Next-to-Minimal Supersymmetric Standard Model (NMSSM) \cite{review}, owing to 
the introduction of an extra complex Higgs  singlet
field, which only couples to the two MSSM-type Higgs doublets, embeds a
Higgs sector which comprises a total of seven mass
eigenstates: a charged pair $h^\pm$, three CP-even Higgses
$h_{1,2,3}$ ($m_{h_1}<m_{h_2}<m_{h_3}$) and two CP-odd
Higgses $a_{1,2}$ ($m_{a_1}<m_{a_2}$). 

Consequently, Higgs phenomenology in the NMSSM can be very different from that of the
MSSM. As a key example, over the past few years, there have been several attempts
to extend the so-called `No-lose theorem' of the MSSM -- stating 
 that at least one MSSM Higgs boson should be observed through
the usual SM-like production and decay channels at the Large Hadron Collider (LHC) 
 throughout the entire MSSM parameter space \cite{NoLoseMSSM} --
to the case of the NMSSM \cite{NMSSM-Points,NoLoseNMSSM1,Shobig2}. From this perspective,
it was realised that at least one NMSSM Higgs boson should remain observable 
at the LHC over the NMSSM parameter space that does not allow any Higgs-to-Higgs 
decay mode. In contrast, when a light non-singlet (and, therefore, potentially visible) CP-even
Higgs boson, $h_1$ or $h_2$, decays mainly to two very light 
CP-odd Higgs bosons, $h_{1,2}\to a_1 a_1$, one
may not have any Higgs signal of statistical significance at the LHC \cite{dirk}. In fact, 
further violations to the theorem may well occur if
one enables Higgs-to-sparticle decays (e.g., into neutralino pairs, yielding invisible Higgs signals)
~\footnote{A sparticle is the Supersymmetry (SUSY) partner of an ordinary particle.}.
While there is no conclusive evidence on whether a `No-lose theorem' can be proved for the NMSSM
at the LHC, there has also been put forward an orthogonal approach: to see when a, so to say,
`More-to-gain theorem' for the LHC \cite{Shobig1,Erice,CPNSH} can be formulated 
within the NMSSM. That is, whether there exist regions
of the NMSSM parameter space where more and/or different Higgs
states are visible at the LHC than those available within the MSSM.
 
In our attempt to overview both such possibilities, we assume here a light CP-odd Higgs boson, $a_1$, emerging 
from the decay $h_2\to Za_1$, where the heavy CP-even state, $h_2$, is produced in association with $b$-quark pairs 
at the LHC. This work complements the one carried out in a previous paper
where the light $a_1$ state is produced in pairs from decays of not only $h_2$ states but also $h_1$ ones, always from 
the aforementioned production mode \cite{Almarashi:2011te}. Direct $a_1$ production in association with $b\bar b$ pairs
was considered in \cite{Almarashi:2010jm} for the case of $a_1\to \tau^+\tau^-$ as well as $a_1\to\gamma\gamma$ decays, in 
\cite{Almarashi:2011hj} for the case of the $a_1\to\mu^+\mu^-$ channel and in \cite{Almarashi:2011bf} for the $a_1\to b\bar b$ mode.

For our study of the NMSSM Higgs sector
we have used NMSSMTools \cite{NMHDECAY,NMSSMTools}, which is a numerical package
computing the masses, couplings and decay widths of all the Higgs
bosons of the NMSSM in terms of its input parameters at the Electro-Weak (EW)
scale. NMSSMTools also takes into account theoretical constraints as well as
experimental limits, including the unconventional channels relevant for the NMSSM. 

Here, instead of postulating unification or taking into account the 
SUSY breaking mechanism, we fix the soft SUSY breaking terms to a very high value, 
so that they give a small or no contribution at all to the outputs of the parameter scans. 
Consequently, we are left with six free parameters at the EW scale, uniquely defining the NMSSM Higgs sector
at tree-level. Our  parameter space is in particular identified through the Yukawa couplings $\lambda$ and
$\kappa$, the soft trilinear terms $A_\lambda$ and $A_\kappa$, plus 
tan$\beta$ (the ratio of the 
Vacuum Expectation Values (VEVs) of the two Higgs doublets) and $\mu_{\rm eff} = \lambda\langle S\rangle$
(where $\langle S\rangle$ is the VEV of the Higgs singlet). 

In order to make a comprehensive study of the NMSSM parameter space,
we have used NMSSMTools to scan over the aforementioned six parameters taken in the following 
intervals:
\begin{center}
$
0.0001 <\lambda< 0.7,~~~0<\kappa<0.65,~~~1.6<\tan\beta<54,
$
\\[0.15cm]
$100~{\rm GeV}~<\mu_{\rm eff}< 1~{\rm TeV},~~~-1~{\rm TeV}~<A_{\lambda}<1~{\rm TeV},
$
\\[0.15cm]
$
-10~{\rm GeV}~< A_{\kappa}<0.
$
\end{center}
Soft terms which are fixed in the scan include:\\
$\bullet\phantom{a}m_{Q_3} = m_{U_3} = m_{D_3} = m_{L_3} = m_{E_3} = 1$ TeV, \\
$\bullet\phantom{a}A_{U_3} = A_{D_3} = A_{E_3} = 1.2$ TeV,\\
$\bullet\phantom{a}m_Q = m_U = m_D = m_L = m_E = 1$ TeV,\\
$\bullet\phantom{a} M_1 = M_2 = M_3 = 1.5$ TeV.\\

We have finally performed our scan over 2$\times$10$^7$ randomly
selected points in the specified parameter space. The points which violate the constraints
(either theoretical or experimental)
are automatically eliminated by NMSSMTools. 

The surviving data points are then used to determine the
cross-sections for NMSSM Higgs hadro-production by using CalcHEP \cite{CalcHEP} for signals
and MadGraph \cite{MadGraph} for backgrounds. As the SUSY mass scales 
have been arbitrarily set well above the EW one (see above), 
the NMSSM Higgs production modes
exploitable in simulations at the LHC are those involving couplings to
heavy ordinary matter only. Amongst the productions channels onset by the latter, 
we focus here on 
\begin{equation}\label{prod}
pp(q\bar q,  gg)\to b\bar b~{h_2},
\end{equation}
i.e., Higgs production in association with a $b$-quark pair, followed by
\begin{equation}\label{dec}
h_2\to Za_1
\end{equation}
(incidentally, notice that $h_1$ is never heavy enough to enable sizable $Z a_1$ decays) and then
\begin{equation}\label{sign}
Za_1\to (jj)(\tau^+\tau^-)~({\rm{where}}~j={\rm{jet}}).
\end{equation}
Notice that we discard here the possibility of $a_1\to \mu^+\mu^-$ decays, for two reasons: on the one hand, 
the mass region below the $\tau^+\tau^-$ threshold is severely constrained (see \cite{Lebedev}
and references therein); on the other hand, the $\mu^+\mu^-$ decay rates are $\approx (m_\mu^{\rm pole}/m_\tau^{\rm pole})^2$ 
times suppressed with respect to the $\tau^+\tau^-$ ones, so that they would be numerically irrelevant (see forthcoming figures). 
As for $a_1\to b\bar b$ decays, while being in turn more numerous than the $\tau^+\tau^-$ ones (above the $2m_b$ threshold) 
by an amount $\propto (m_b(m_{h_2})/m_\tau^{\rm pole})^2$, they are overwhelmed by QCD backgrounds in $Z$~+jet final states. 
Notice that the $a_1$ masses of relevance to this study will typically range between 15 and 60 GeV or so.   

We adopt herein CTEQ6L \cite{cteq} as parton distribution functions, with scale $Q=\sqrt{\hat{s}}$, 
the centre-of-mass energy
at parton level, for all processes computed. We finally assume $\sqrt s=14$ TeV throughout for the 
LHC energy. Also, in our numerical 
analyses, we have taken $m_b(m_b)=4.214$ GeV and $m_t^{\rm pole}=171.4$ GeV for the (running) bottom- and 
(pole) top-quark mass, 
respectively, while we have input $m_\tau^{\rm pole}=1.777$ GeV and $m_\mu^{\rm pole}=0.1057$ GeV for the (pole) tau- and 
(pole) muon-lepton mass, respectively. 

As an initial step towards the analysis of the data, we have computed the production cross-section
$\sigma(pp\to b\bar b h_2)$ times the decay BR$(h_2\to Za_1)$  against 
the BR itself, see Fig.~1.
%
%
\begin{figure}
 \centering\begin{tabular}{c}
  \includegraphics[scale=0.65]{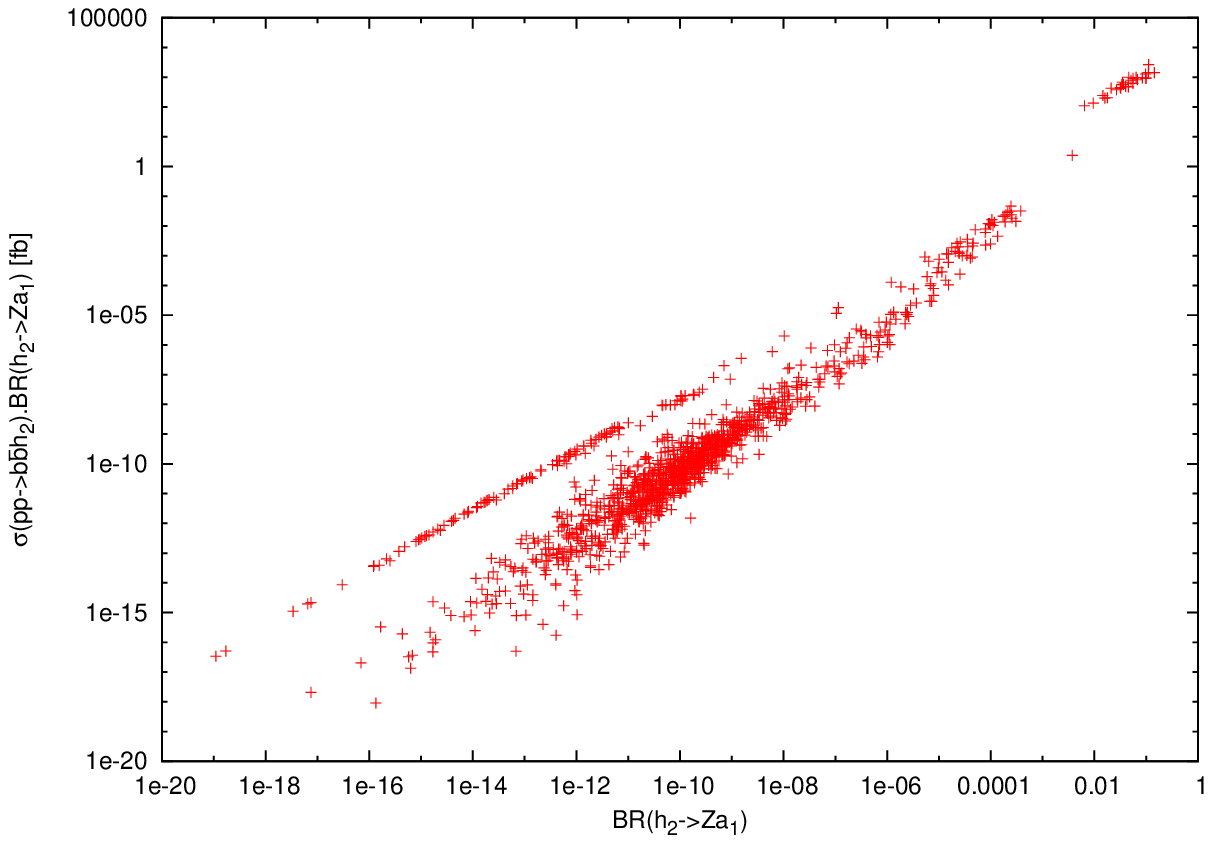}
 \end{tabular}
\label{fig:BR}
\caption{$\sigma(pp\to b\bar b h_2)$ BR$(h_2\to Za_1)$ vs.  
BR$(h_2\to Za_1)$.}
\end{figure}
It is clear from this plot that most of the NMSSM parameter space yields rather small rates: in fact, only when
the BR$(h_2\to Za_1)$  is large the overall process (\ref{prod})--(\ref{dec}) can offer some chances of detection. 
In this scatter plot, one can notice a population of points with cross-section of up to several pb's. This region
of parameter space is the only one exploitable at the LHC and reflects a well defined setup for the six input parameters of the NMSSM. Generally, such
points correspond to generic and comfortably perturbative $\lambda$ and $\kappa$ (between 0.05 and 0.15), 
very large $\tan\beta$ (above 35),
rather large (and positive) $\mu_{\rm eff}$ and large (and negative) $A_\lambda$ (both between 200 and 900 GeV)
and slightly negative $A_{\kappa}$ (between $-10$ and $-1$ GeV).  

The rates in the previous figure refer to on-shell $Z$ and $a_1$ though. One should clearly extract these from suitable decays,
inevitably coming with a decay probability less than 1.
Here, as intimated earlier on, we attempt the case of hadronic decays of the gauge boson and $\tau$-decays of the CP-odd Higgs boson. 
The corresponding signal yield for the ensuing final state is displayed in Fig.~2, now mapped against $m_{h_2}$.  
\begin{figure}
 \centering\begin{tabular}{c}
  \includegraphics[scale=0.65]{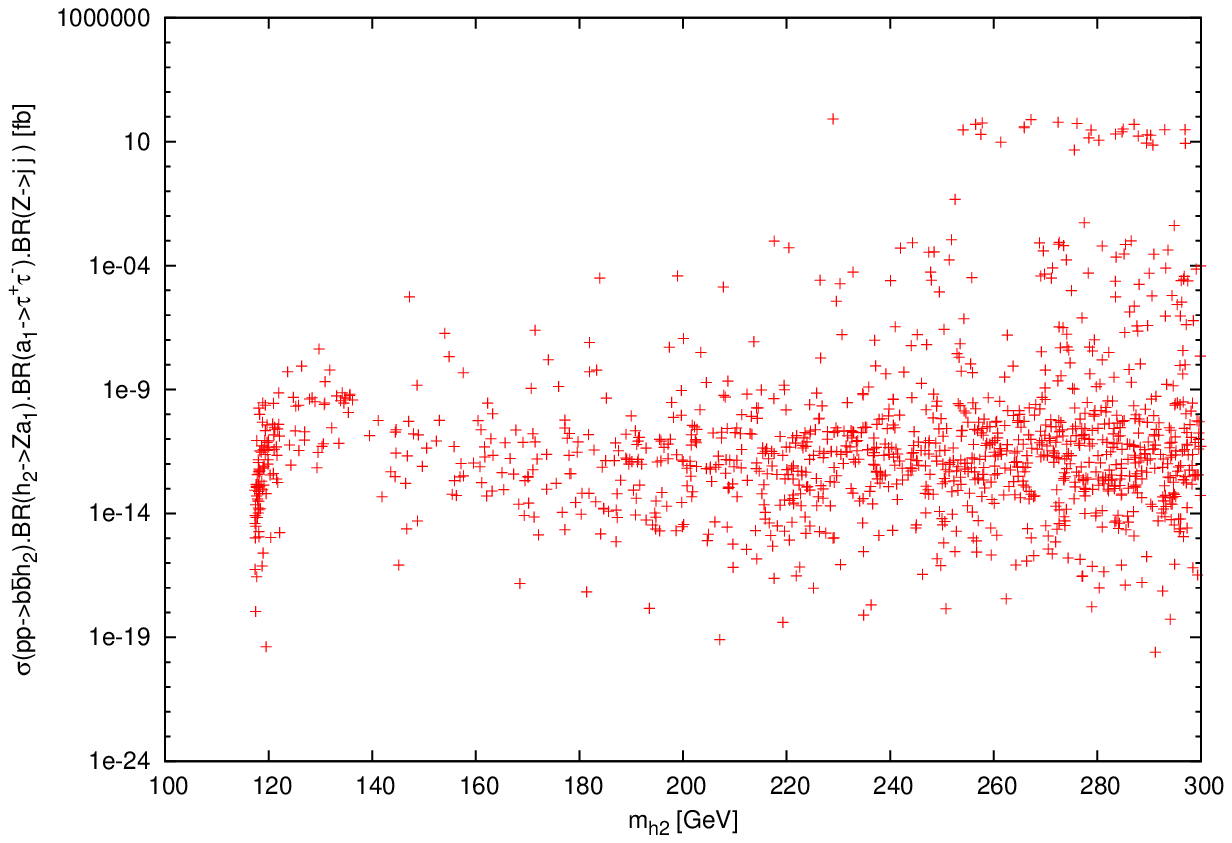}
 \end{tabular}
\label{fig:signature}
\caption{$\sigma(pp\to b\bar b h_2)$ BR$(h_2\to Za_1)$ BR$(a_1\to\tau^+\tau^-)$ 
BR$(Z\to jj)$ vs $m_{h_2}$.}
\vspace*{-6.35cm}\hspace*{-0.5cm}{benchmark points $\longrightarrow$}\vspace*{6.0cm}\hspace*{0.5cm}
\end{figure}
After folding in the above decay probabilities, one is left with event rates at the ${\cal O}$(100) fb level at the most.
While clearly this number is not very large, signal events may still be detectable at planned LHC luminosities, especially
if the background can be successfully reduced to manageable levels. Before proceeding to do so, it is worth mentioning that 
this kind of event rates are only found for $h_2$ masses well above 250 GeV. 

We perform next a partonic signal-to-background ($S/B$) analysis. Amongst the backgrounds, we consider here 
what we verified to be
the dominant one, the irreducible noise induced by $pp\to b\bar b Z \tau^+\tau^-$ channels (amounting
to 60 Feynman diagrams in the unitary gauge). As benchmarks for the $S/B$ analysis 
we have chosen four points in the above parameter space region (see Fig.~2), as illustrative examples. These benchmarks
are given in Tab.~I.

\begin{table}\label{tab:benchmarks}
\begin{flushleft}
\begin{tabular}{|c||c|c|c|}
\hline
Point & $\lambda$ & $\kappa$ & tan$\beta$    \\ \hline
1 &
0.11784333E+00 & 0.99759129E-01 & 0.45413385E+02\\ \hline
2 &
0.36954734E-01 & 0.74106016E-01 & 0.35731787E+02\\ \hline
3 &
0.55822718E-01 & 0.41921611E-01 & 0.47269331E+02\\ \hline
4 &
0.15630654E+00 & 0.84945744E-01 & 0.48122165E+02\\ \hline
\end{tabular}
\begin{tabular}{|c||c|c|c|}
\hline
Point & $\mu_{\rm eff}$ [GeV] & $A_\lambda$ [GeV] & $A_\kappa$ [GeV] \\ \hline
1 &
0.65363068E+03 & $-0.56108302$E+03 & $-0.97511471$E+01 \\ \hline
2 &
0.33852788E+03 & $-0.68364824$E+03 & $-0.13135077$E+01 \\ \hline
3 &
0.38990121E+03 & $-0.29909449$E+03 & $-0.53814631$E+01 \\ \hline
4 &
0.42293377E+03 & $-0.23627361$E+03 & $-0.74575765$E+01 \\ \hline
\end{tabular}
\begin{tabular}{|c||c|c|c|c|}
\hline
Point & $m_{a_1}$ [GeV] & $m_{h_2}$ [GeV] & $\Gamma_{a_1}$ [GeV] & $\Gamma_{h_2}$ [GeV] \\ \hline
1 &
44.1 & 272.4 & 0.205 & 6.57    \\ \hline
2 &
15.9 & 261.4 & 0.01879 & 4.972 \\ \hline
3 &
61.5 & 275.6 & 0.00293 & 7.86  \\ \hline
4 &
31.9 & 287.9 & 0.0674 & 8.37   \\ \hline
\end{tabular}
\begin{tabular}{|c||c|c|}
\hline
Point & $\sigma(pp\to b\bar bh_2)$ BR$(h_2\to Za_1)$ [fb] & BR$(a_1\to \tau^+\tau^-)$  \\ \hline
1 & 1286.35 & 0.141549576 \\ \hline
2 &   377.3 & 0.07615127  \\ \hline
3 &   109.0 & 0.125658544 \\ \hline
4 &   447.1 & 0.111203597 \\ \hline
\end{tabular}
\caption{The NMSSM benchmark points used in the $S/B$ analysis. }
\end{flushleft}
\end{table}
In the light of the cross-section and decay rates in the table, and
recalling that BR$(Z\to jj)\approx70\%$, the total inclusive cross-section for
the signal in (\ref{prod})--(\ref{sign}) for point 1(2)[3]\{4\} is
127(20)[10]\{35\} fb. The SM irreducible background inclusive rate is 7.6 fb. 
After implementing the following acceptance cuts~\footnote{Here, for the sake of illustration,
we take the $\tau$'s on-shell.}
$$\Delta R (i,j)>0.4~(i,j=b,\bar b, j, j,\tau^+\tau^-),$$
$$\arrowvert\eta(i)\arrowvert <2.5~(i=b,\bar b, j, j,\tau^+\tau^-),$$
\begin{equation}
p_{T}(i)>15~{\rm{GeV}}~(i=b,\bar b, j, j, \tau^+, \tau^-),
\label{acceptance-cuts}
\end{equation}
and the selection ones as well
\begin{equation}
|M_{jj}-M_Z|<15~{\rm GeV},~~~
|M_{\tau^+\tau^-}-m_{a_1}|<15~{\rm GeV}
\label{selection-cuts}
\end{equation}
(wherein $m_{a_1}$ need not be known beforehand, as it would be seen in the 
reconstructed $M_{\tau^+\tau^-}$ distribution:
here it assumes the values of Tab.~I), 
we obtain for the signal cross-sections the values
2.27(0.12)[0.21]\{0.65\} fb for point 1(2)[3]\{4\} 
so that the typical signal efficiency is approximately 
1.78(0.57)[2.21]\{1.87\}\%, respectively~\footnote{This
includes a factor $\varepsilon_b$ for each of the two $b$-tags which would be required 
(typically, $\varepsilon=60\%$ for the above cut in $p_T(b,\bar b)$)
to isolate the signal we have discussed. Notice that to require to tag both $b$'s in 
the final state with 
transverse momentum above 15 GeV induces a reduction of the signal cross sections, for
the $m_{h_2}$ values considered here, of about 7\%, see Fig. 9 of
\cite{Almarashi:2011te}.}). The selection efficiency strongly depends on the $m_{a_1}$ value, see Fig.~3,
in the sense that the lighter $a_1$ the softer its decay products (the $\tau^+\tau^-$ pair), so that the
$p_T(\tau^+,\tau^-)$ selects fewer signal events: recall in fact that one has 
$m_{a_1}=44.1(15.9)[61.5]\{31.9\}$ GeV for point 1(2)[3]\{4\}.   
Notice that after all such constraints the SM irreducible background is essentially removed
altogether, as its rate goes down dramatically, to 
${\cal O}(10^{-4})$ fb for all points 1--4.
\begin{figure}
 \centering\begin{tabular}{c}
  \includegraphics[scale=0.40,angle=90]{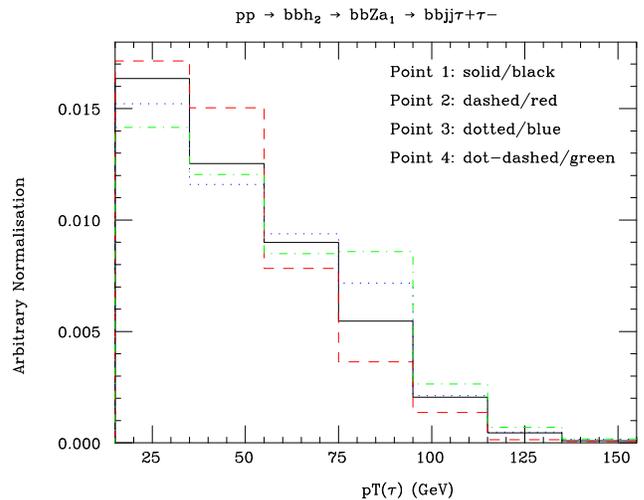}
 \end{tabular}
\label{fig:pT}
\caption{The (reconstructed) $\tau$ transverse momentum distribution
for the four benchmark points in Tab.~I.}
\end{figure}
The distributions in invariant mass $m_{jj\tau^+\tau^-}$, that one could obtain after reconstructing
the $\tau$ decay products, for the four benchmark points, are given in Fig.~4, showing the Breit-Wigner peaks
signaling the $h_2\to Za_1$ resonances. Integrating over the 
entire spectrum, one obtains 2274(115)[213]\{653\} signal events for point 1(2)[3]\{4\}, essentially background 
free, assuming 1000 fb$^{-1}$ of integrated luminosity.  
\begin{figure}
 \centering\begin{tabular}{c}
  \includegraphics[scale=0.40,angle=90]{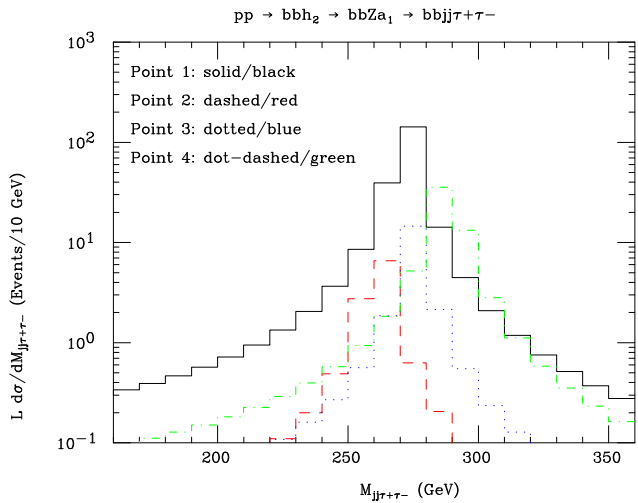}
 \end{tabular}
\label{fig:Mass}
\caption{The (reconstructed) $h_2$ mass peaks in terms of the $jj\tau^+\tau^-$ invariant mass distribution
for the four benchmark points in Tab.~I. We plot the number of events after an integrated luminosity
L~=~1000 fb$^{-1}$.}
\end{figure}

In summary, 
we have proven that there exists a small but well defined region of parameter space where the $h_2$ and $a_1$ states of the NMSSM,
both with a mixed singlet and doublet nature, could potentially be detected at the LHC if 250 GeV $\lsim m_{h_2}\lsim$ 300 GeV and 
$15$ GeV $\lsim m_{a_1}\lsim 60$ GeV, in the 
$h_2\to Za_1\to jj\tau^+\tau^-$ mode, when  the CP-even Higgs state is produced in association with a $b\bar b$ pair for rather large 
tan$\beta$, large (and positive) $\mu_{\rm eff}$, large (and negative) $A_\lambda$,
and slightly negative $A_{\kappa}$, for typically perturbative values of $\lambda$ and $\kappa$.
 After a realistic $S/B$ analysis at parton level, we have in fact produced results showing that
the extraction of such a signal above the dominant irreducible SM background should be feasible using standard
reconstruction techniques \cite{ATLAS-TDR,CMS-TDR} and large LHC luminosities, i.e., after several years
of running at design values or rather promptly  at the Super-LHC \cite{SLHC}. While more refined
analyses, incorporating $\tau$-decays, parton shower, hadronisation and detector effects, are needed in order to delineate the 
true discovery potential of the LHC
over the actual NMSSM parameter space, we are confident that our results are a step in the right direction to both:
(i) prove the existence of a `More-to-gain theorem' at the CERN collider for the NMSSM with respect to the MSSM
(as Higgs~$\to Z$~Higgs' signals are only possible in the latter scenario in parameter space regions
already excluded by experimental data) and (ii) 
to establish a `No-lose theorem' for the NMSSM at the LHC (as some of the parameter regions where the aforementioned signal can be detected overlap 
with those where  $h_{1,2}\to a_1a_1$ decays might be ineffective in extracting an NMSSM Higgs signal).
\vspace*{-0.25cm}
\section*{Acknowledgments}
\vspace*{-0.25cm}
This work is 
supported in part by the NExT Institute. M. M. A. acknowledges
funding from Taibah University (Saudi Arabia).

\newpage

\end{document}